# Dissipative dynamics of qubits driven by a bichromatic field in the dispersive regime

A.P. Saiko[1], R. Fedaruk[2], A. Kolasa[2], S.A. Markevich[1]

[1]Scientific-Practical Materials Research Centre NAS of Belarus, Minsk, Belarus

[2]Institute of Physics, University of Szczecin, 70-451, Szczecin, Poland

E-mail: saiko@ifttp.bas-net.by; fedaruk@wmf.univ.szczecin.pl

We study the coherent dynamics of relaxing qubits driven by a bichromatic radiation in the dispersive regime, when detuning of the frequency $\omega_{rf}$ of a longitudinal radiofrequency field from the Rabi frequency $\omega_1$ in a transverse microwave field is comparable in magnitude to $\omega_{rf}$ and $\omega_1$. We analytically describe this regime beyond the rotating wave approximation and find that the dominant feature of dynamics of qubits is the shift of the Rabi frequency caused by the dynamical Zeeman and Bloch-Siegert-like effects. These fundamental effects can be experimentally separated because, unlike the Bloch-Siegert effect, the dynamical Zeeman effect depends on the detuning sign. Our theoretical results are in a good agreement with the experimental data obtained in pulse EPR for the $E_1'$ centers in crystalline quartz.

## 1. Introduction

It is known that the resonant interaction between electromagnetic field and a two-level quantum system (qubit) induces Rabi oscillations.[1] The Rabi frequency $\omega_1$ is defined by the amplitude of the electromagnetic field and usually is much smaller than the energy difference (in frequency units) between the qubit's states. Studies of spectroscopic and dissipative properties of qubits based on this phenomenon are often termed the Rabi spectroscopy and are used for a wide range of physical objects, including, among others, nuclear and electron spins in NMR and EPR, quantum dots, flux, and charge qubits in superconducting systems. The second electromagnetic field with the frequency close to $\omega_1$ can strongly change Rabi oscillations giving rise to the so-called Rabi (nutational) resonance. The Rabi resonance in the bichromatic field was observed in NMR (rotary saturation[2,3]), EPR,[4-6] and optical[7] experiments employing both the continuous-wave and coherent time-resolved techniques . At the Rabi resonance, due to the multiplication of qubits, additional Rabi oscillations with the frequency defined by the amplitude of the second field are observed.[8] The application of the Rabi resonance to the readout of qubit states in superconducting and spin systems is considered in Refs. 9-11. In all these investigations qubits were driven by two classical fields with strongly different frequencies.

On the other hand, of practical interest is the dispersive limit[12-15] of the interaction between a qubit and quantum electromagnetic field in a resonator. In this regime, detuning of the field frequency $\omega$ from the quantum transition frequency $\omega_0$ is comparable in magnitude to both $\omega_0$ and $\omega$ itself. A non-destructive measurement of the qubit state can be performed by probing the resonator. The Jaynes-Cummings model is usually used in such studies.[12-14] In the most interesting cases, the interaction between the qubits and the quantized field is strong. In order to describe this situation taking into account energy dissipation and dephasing of qubit states, numerical methods are commonly used. Therefore, it is important to develop analytically solvable models that include dissipative processes in the Rabi spectroscopy. In the present paper, we consider the dispersive regime for spin qubits driven by a classical bichromatic field. In this case, qubit states are «dressed» by the high-frequency field with the frequency $\omega_{mw}$ and the Rabi oscillations with the frequency $\omega_1$ are induced. The second low-frequency field with the frequency $\omega_{rf}$ weakly modulates the frequency of qubit transition $\omega_0$, and the dispersive condition $|\omega_{rf} - \omega_1| \sim \omega_{rf}, \omega_1$ is fulfilled. As far as we know, such a situation has not been studied until now. In this case, analytical solutions adequately describing dissipative dynamics of qubits beyond the rotating wave approximation can be obtained. Here, dissipative processes are described by the Liouville equation introducing the relaxation operator in its usual form. In Sec. 2, we show that coherent dynamics of the system is caused by the dynamical Zeeman and Bloch-Siegert-like effects, and dissipative processes can be characterized by the renormalized longitudinal and transverse relaxation rates depending on $\omega - \omega_0$ and $\omega_1$. In Sec. 3, we present experimental results confirming our theoretical predictions.

## 2. Theory

Let a spin qubit be in a microwave (MW) field directed along the *x* axis of the laboratory frame, in a radiofrequency (RF) field and a static magnetic field, both directed along the *z* axis. The Hamiltonian of the qubit in these fields can be written in the form:

$$H = H_0 + H_\perp(t) + H_\parallel(t), \qquad (1)$$

where $H_0 = \omega_0 s^z$ is the Hamiltonian of the Zeeman energy of the qubit in the static magnetic field $B_0$, $\omega_0 = \gamma B_0$, $\gamma$ is the gyromagnetic ratio; $H_\perp(t) = \omega_1(s^+ + s^-)\cos\omega_{mw}t$ and $H_\parallel(t) = 2\omega_2 s^z \cos(\omega_{rf} t + \psi)$ are the Hamiltonians of the qubit interaction with linearly polarized MW and RF fields, respectively. Here $B_1$ and $B_2$, $\omega_{mw}$ and $\omega_{rf}$ are the respective amplitudes and frequencies of the MW and RF fields, $\psi$ is the phase of the RF field, the MW phase is fixed and equals zero, and the counter-rotating component of the MW field is neglected. Moreover, $\omega_1 = \gamma B_1$ and $\omega_2 = \gamma B_2$ denote the respective interaction constants, whereas $s^{\pm,z}$ are components of the spin operator, describing the state of the qubit and satisfying the commutation relation: $[s^+, s^-] = 2s^z$, $[s^z, s^\pm] = \pm s^\pm$.

Dynamics of the system with the Hamiltonian (1) including relaxation processes is described by the Liouville equation for the density matrix $\rho$:

$$i\hbar \frac{\partial \rho}{\partial t} = [H, \rho] + i\Lambda\rho \qquad (2)$$

(further we assume $\hbar = 1$). The relaxation operator $\Lambda$, which describes the energy dissipation and the dephasing of the qubit, can be written as follows:

$$\Lambda\rho = \frac{\gamma_{21}}{2}(2s^-\rho s^+ - s^+s^-\rho - \rho s^+s^-) +$$

$$+ \frac{\gamma_{12}}{2}(2s^+\rho s^- - s^-s^+\rho - \rho s^-s^+) + \frac{\eta}{2}(2s^z\rho s^z - \frac{1}{2}\rho), \qquad (3)$$

where $\gamma_{21}$ and $\gamma_{12}$ are the rates of the transitions from the excited state 2 of the qubit to its ground state 1 and vice versa, and $\eta$ is the dephasing rate. After three canonical transformations $\rho_3 = u_3^+ u_2^+ u_1^+ \rho u_1 u_2 u_3$, where $u_1 = \exp(-i\omega_{mw} t s^z)$, $u_2 = \exp(-i\theta s^y)$, $u_3 = \exp(-i\Omega_R t s^z)$, equation (2) is transformed into $i\partial\rho_3/\partial t = [H_3, \rho_3] + i\Lambda_3\rho_3$. Here

$$H_3 = [\frac{2\omega_2\Delta}{\Omega_R}s^z - \frac{\omega_1\omega_2}{\Omega_R}(s^+ e^{i\Omega_R t} + s^- e^{-i\Omega_R t})]\cos(\omega_{rf} t + \psi), \qquad (4)$$

$$\Lambda_3\rho_3 = \frac{\Gamma_\downarrow}{2}(2s^-\rho_3 s^+ - s^+s^-\rho_3 - \rho_3 s^+s^-) +$$

$$+ \frac{\Gamma_\uparrow}{2}(2s^+\rho_3 s^- - s^-s^+\rho_3 - \rho_3 s^-s^+) + \frac{\Gamma_\varphi}{2}(2s^z\rho_3 s^z - \frac{1}{2}\rho_3), \qquad (5)$$

$$\Gamma_\downarrow = \frac{1}{4}(\gamma_{21} + \gamma_{12})(1 + \cos^2\theta) + \frac{1}{2}(\gamma_{21} - \gamma_{12})\cos\theta + \frac{1}{4}\eta\sin^2\theta, \qquad (6)$$

$\Gamma_\uparrow = \frac{1}{4}(\gamma_{21} + \gamma_{12})(1 + \cos^2\theta) - \frac{1}{2}(\gamma_{21} - \gamma_{12})\cos\theta + \frac{1}{4}\eta\sin^2\theta$, $\Gamma_\varphi = \eta\cos^2\theta + (\gamma_{21} + \gamma_{12})\sin^2\theta$,

$\Omega_R = (\omega_1^2 + \Delta^2)^{1/2}$ is the generalised Rabi frequency in the MW field, $\Delta = \omega_0 - \omega_{mw}$, $\cos\theta = \Delta/\Omega_R$, $\sin\theta = \omega_1/\Omega_R$. In order to obtain the relaxation operator (5), we assume that the conditions $\gamma_{21}, \gamma_{12}, \eta \ll \omega_{rf}, \omega_1$ are fulfilled, and the non-diagonal terms in the operator (5) that contain the products of spin operator pairs $s^\pm$ and $s^z$, $s^+$ and $s^+$, $s^-$ and $s^-$ can be neglected. Therefore, after the unitary transformations, the form of the operator (3) remains unchanged, but the dissipation and phase relaxation rates (6) are changed.

In the dispersive regime,

$$\omega_2, |\omega_{rf} - \omega_1| \sim \omega_{rf}, \omega_1, \qquad (7)$$

rapidly oscillating terms in the transformed Liouville equation can be eliminated by the Krylov–Bogoliubov–Mitropolsky method (see, e.g. Refs. 8, 16), and we obtain this equation in the following form: $i\partial\langle\rho_3\rangle/\partial t = \left[H_{eff},\langle\rho_3\rangle\right] + i\langle\Lambda_3\rangle\langle\rho_3\rangle$, where symbol $\langle...\rangle$ denotes time averaging over rapid oscillations of the type $\exp(\pm i\omega_{rf}t)$, $\exp(\pm i(\omega_{rf}-\Omega_R)t)$, $\exp(\pm i(\omega_{rf}+\Omega_R)t)$,

$$H_{eff} = \frac{i}{2}\langle[\int^t d\tau(H_3(\tau)-\langle H_3(\tau)\rangle), H_3(t)]\rangle = \delta\Omega s^z. \tag{8}$$

Square brackets in the definition $H_{eff}$ denote the commutation operation. The averaging does not affect the form of the relaxation operator, $\langle\Lambda_3\rangle = \Lambda_3$. The expression

$$\delta\Omega = \frac{\omega_1^2\omega_2^2}{2\Omega_R^2}\left(\frac{1}{\Omega_R-\omega_{rf}}+\frac{1}{\Omega_R+\omega_{rf}}\right) \equiv \delta\Omega_Z + \delta\Omega_{B-S} \tag{9}$$

describes the frequency shift in the Rabi oscillations. The first term $\delta\Omega_Z$ in Eq. (9) is caused by the dynamical (ac) Zeeman effect[17] which is due to taking into account the rapid oscillations $\exp(\pm i(\omega_{rf}-\Omega_R)t)$ in the second order of the nonsecular perturbation theory. The second term $\delta\Omega_{B-S}$ describes the Bloch–Siegert-like shift[18] originating from the antiresonance oscillations $\exp(\pm i(\omega_{rf}+\Omega_R)t)$ in the same approximation. Since $H_{eff}$ and $\Lambda_3$ have a diagonal form, the density matrix in the laboratory frame is written as: $\rho(t) = u_1 u_2 u_3 \exp(-i\delta\Omega s^z t)((\exp(\Lambda_3 t)(u_2^+\rho(0)u_2))\exp(i\delta\Omega s^z t)u_3^+ u_2^+ u_1^+$. Here $\rho(0) = 1/2 - s^z$ if at the initial moment the qubit is found in the ground state. Disentangling $\rho(t)$, we obtain the absorption signal $V$ and the population difference $W$ of the qubit levels in the rotating frame which rotates with frequency $\omega_{mw}$ around the $z$ axis of the laboratory frame:

$$V(t) = \frac{1}{2i}(\langle 1|\rho(t)|2\rangle - \langle 2|\rho(t)|1\rangle) = \frac{1}{2}\sin\theta\exp(-\Gamma_\perp t)\sin(\Omega_R + \delta\Omega)t, \tag{10}$$

$$W(t) = \frac{1}{2}(\langle 2|\rho(t)|2\rangle - \langle 1|\rho(t)|1\rangle) = -\frac{1}{2}[\sin^2\theta\exp(-\Gamma_\perp t)\cos(\Omega_R + \delta\Omega)t +$$
$$+\cos^2\theta\exp(-\Gamma_\| t) + \frac{\Gamma_\downarrow - \Gamma_\uparrow}{\Gamma_\|}(1-\exp(-\Gamma_\| t)\cos\theta], \tag{11}$$

where $|1\rangle$ and $|2\rangle$ are the ket-vectors of the ground and excited states of the qubit, $\Gamma_\| = \Gamma_\downarrow + \Gamma_\uparrow$, $\Gamma_\perp = (\Gamma_\downarrow + \Gamma_\uparrow + \Gamma_\varphi)/2$. The renormalized relaxation rates $\Gamma_\perp$ and $\Gamma_\|$ can be expressed in terms of the longitudinal $T_1$ and transverse $T_2$ relaxation times in the laboratory frame: $\Gamma_\perp = (1/2T_2)(1+\cos^2\theta) + (1/2T_1)\sin^2\theta$, $\Gamma_\| = (1/T_2)(1+\sin^2\theta) + (1/T_1)\cos^2\theta$. Note

that the phase of the RF field does not influence the dissipative dynamics of the qubit, because there is no $\psi$ in the expressions for $V$ and $W$. It follows from Eq. (11) that at the initial moment of $t = 0$ the qubit is in the ground state: $W(0) = -1/2$, and when $t \to \infty$, it enters the stationary regime: $W(\infty) = -((\Gamma_\downarrow - \Gamma_\uparrow)/2\Gamma_\parallel)\cos\theta$ (in the rotating frame).

According to Eqs. (10) and (11), the bichromaric field in the dispersive regime shifts the Rabi frequency $\Omega_R$ by the value $\delta\Omega$ given by equation (9). At the exact resonance of the qubit with the MW field ($\Delta = 0$), the absorption signal and the population difference oscillate with the effective Rabi frequency

$$\Omega_R^{eff} = \omega_1\left(1 + \frac{\omega_2^2}{\omega_1^2 - \omega_{rf}^2}\right), \qquad (12)$$

and the frequency shift $\delta\Omega(\Delta = 0) \equiv \delta\Omega^0 = \omega_1\omega_2^2/(\omega_1^2 - \omega_{rf}^2)$. Without the RF field ($\omega_2 = 0$), there is no the frequency shift and the Rabi oscillation frequency of the absorption signal and the population difference equals $\omega_1$.

In condensed matter, frequencies of quantum transitions of qubits can be different, and their distribution is usually described by the Gauss function $g(\Delta)$. Therefore, equations (10) and (11) should be averaged over the inhomogeneous distribution of the frequencies $\Delta$, for example: $\{V(t)\}_{Av} = \int_{-\infty}^{+\infty} d\Delta g(\Delta)V(\Delta,t)$. If the width of the inhomogeneous line is much larger than $\omega_1$, it may be considered as an infinite one. Neglecting the dependences of $\Gamma_\perp$ and $\delta\Omega$ on $\Delta$, we obtain:

$$\{V(t)\}_{Av} = \pi g(0)\omega_1 \exp(-\Gamma_\perp^0 t)[J_0(\omega_1 t)\cos(\delta\Omega^0 t) - N_0(\omega_1 t)\sin(\delta\Omega^0 t)], \qquad (13)$$

where $J_0(x)$ and $N_0(x)$ are the zero order Bessel function of the first and second kind, respectively, and $\Gamma_\perp^0 = \Gamma_\perp(\Delta = 0)$.

### 3. Experimental results and discussion

The usual Rabi oscillations were induced after the abrupt establishment of resonant interaction between the continuous MW field and the spin system by the pulse of a longitudinal magnetic field.[19] Initially, the spin system was exposed to static magnetic field $B$ at which no absorption of MW radiation occurred. Then, the magnetic field was jumpwise (within about 120 ns) changed to the resonant value $B_0$. Due to the Zeeman effect, the resonant frequency of the spin system changed with a jump in the magnetic field to $\omega_0 = \gamma B_0$ and became equal to the frequency $\omega_{mw}$ of the MW field. During the action of the magnetic pulse, the resonant interaction of the MW field with the spin system induced the Rabi oscillations. At the resonant excitation ($\omega_{mw} = \omega_0$) the frequency of the Rabi oscillation observed in the EPR absorption signal equals $\omega_1$.[19] Pulses of a magnetic field with amplitude $\Delta B = B - B_0$ = 0.12 mT and duration of 10 μs were used. In order to study the dispersive regime in the bichromatic field, in addition to the MW field, a continuous linearly polarized RF field oriented along the static magnetic field was applied.

In the method based on magnetic field pulses, the Rabi oscillations are detected directly during the bichromatic irradiation. This method is instrumentally less demanding with respect to the MW circuit of the EPR spectrometer. The MW circuit of a conventional continuous wave EPR spectrometer can be easily adapted for such experiments. In that case, the special construction of the resonator is used to realize the magnetic field pulses.

The experiments were performed at room temperature using the X-band home-made EPR spectrometer.[19] A homodyne MW bridge with a Gunn diode generator at frequency of 10 GHz and microwave power of up to 250 mW was used. The rectangular $TE_{102}$ resonator was modified by inserting a pair of special metallic rods to form the modulation loop and a lumped capacitance. Due to the lumped capacitance between the metallic rods and the cavity walls, the resonator dimensions and its quality factor became smaller, but the amplitude and the homogeneity of the MW field were increased. A magnetic field pulse and a continuous RF field were produced by passing appropriate currents through the modulation loop inside the measuring resonator. The RF field amplitude was calibrated with accuracy of about 2%. Multichannel digital summation of signals was used to improve the signal-to-noise ratio. The measurements were carried out without the phase synchronization of the RF field with respect to the magnetic-field pulse.

Due to the long relaxation times and the small EPR line width, the $E_1'$ centers ($S$ = 1/2) in neutron-irradiated quartz[6] were used in our experiments as a model two-level spin system. The EPR spectrum of the $E_1'$ centers consists of an isolated line with the width of $\Delta B_{pp}$ = 16 μT at the magnetic field direction along the crystal optical axis.

Fig. 1 (a) shows the absorption signals of the EPR Rabi oscillations of the $E_1'$ centers detected for a fixed MW field amplitude ($\omega_1 / 2\pi$ = 1.03 MHz) without the RF field and in the dispersive regime for two frequencies of the RF field. The observed signals clearly demonstrate the frequency shift of the Rabi oscillations and its dependence on the detuning sign. In accordance with theoretical predictions, the Rabi frequency in the bichromatic field increases for $\omega_{rf} < \omega_1$ and it decreases for $\omega_{rf} > \omega_1$. The Rabi oscillations in both the MW field ($\omega_2 = 0$) and the bichromatic field are observed to decay at the same rate. Our analytical model [Eq. (13)] shows excellent agreement with these observations. Fig. 1 (b) illustrates the results of the calculation using Eq. (13) for $T_2 = 3.5$ μs. Because $T_2$  $T_1 = 0.2$ ms, [20] the influence of longitudinal relaxation is neglected in the calculation. Inspection of Fig. 1 reveals that our theoretical predictions agree well with the experimentally observed dissipative dynamics of qubits driven by the MW and RF fields in the dispersive regime.

The change in the effective Rabi frequency as a function of detuning for the different values of the RF field amplitude at the fixed MW field amplitude ($\omega_1/2\pi = 1.03$ MHz) is presented in Fig. 2. The solid lines in the figure are the theoretical dependencies given by Eq. (12) for the parameters of the bichromatic field used in our experiment. The experimental results are in very good agreement with theoretical predictions. Note that for the RF frequency $\omega_{rf}$ close to $\omega_1$, the analytically obtained Rabi frequency $\Omega_R^{eff}$ diverges. This nonphysical result is due to the violation of the dispersive conditions (Eq. 7).

According to our model, coherent dynamics of the qubits driven by the bichromatic field in the dispersive regime is caused simultaneously by the dynamical Zeeman and Bloch-Siegert effects. This fact is illustrated in Fig. 3 for $\omega_1/2\pi = 1.03$ MHz and $\omega_2/2\pi = 0.24$ MHz. The contribution of each of these effects to the observed frequency shift $\delta\Omega^0$ of the Rabi oscillations is different. Unlike the Bloch-Siegert effect, the dynamical Zeeman effect depends on the detuning sign. At the same time, the Bloch-Siegert effect results in the asymmetric dependence of the shift $\delta\Omega^0$ on the frequency $\omega_{rf}$ of the RF field. Two measurements of the shift $\delta\Omega$ are sufficient to separate the contributions of the dynamical Zeeman and Bloch-Siegert effects. For $\omega_{rf} = a$ and $\omega_1 = b$ satisfying the condition (7), the first measurement of $\delta\Omega$ is performed. In the second measurement, $\omega_{rf} = b$ and $\omega_1 = a$ are used. Here, we assume $\Delta = 0$. Now, we obtain: $(\delta\Omega^0)_{exp1} + (\delta\Omega^0)_{exp2} = \omega_2^2/(a+b) = 2\delta\Omega_{B-S}$, $(\delta\Omega^0)_{exp1} - (\delta\Omega^0)_{exp2} = \omega_2^2/(a-b) = 2\delta\Omega_Z$. For example, for $a/2\pi = 1.2$ MHz, $b/2\pi = 0.6$ MHz and $\omega_2/2\pi = 0.24$ MHz the experimentally obtained values are $\delta\Omega_Z^0 = 0.05$ MHz and $\delta\Omega_{B-S}^0 = 0.02$ MHz. These values agree well with those calculated theoretically: 0.048 MHz and 0.016 MHz.

In our experiments, the Rabi oscillations were observed in the rotating frame in which the effective Rabi frequency could be measured. In this frame, the qubit evolution is described by the density matrix $\rho^I(t) = u_1^+ \rho(t) u_1$. The measurements can be performed in the doubly rotating frame where the qubit evolution is given by the density matrix $\rho^{II}(t) = u_3^+ u_2^+ \rho^I(t) u_2 u_3$. Fig. 4. shows the time evolution of the population difference $W(t)$ of the qubit system in the singly and doubly rotating frames. Equation (11) was averaged over the inhomogeneous distribution of frequencies with the Gauss function $g(\Delta)$. The gray line represents the signal $W(t)$ obtained without the RF field when the frequency shift $\delta\Omega$ vanishes. One can see that the transition into the doubly rotating frame separates the oscillations with the frequency $\Omega_R$. Consequently, only the damped oscillations with the frequency $\delta\Omega$ exist. Without the RF field in the doubly rotating frame, only the relaxation process can be observed (gray line in Fig.4 (b)). Thus, experiments in the doubly rotating frame allow one to observe the frequency shift $\delta\Omega$ directly.

## 4. Summary

We have theoretically and experimentally investigated the interaction of qubits with the bichromatic (MW and RF) field in the dispersive regime when the condition (7) was enforced by choosing the field frequencies ($\omega_{mw}$, $\omega_{rf}$) and the qubit-field couplings ($\omega_1$, $\omega_2$). In this regime beyond the rotating wave approximation, we have found that coherent dynamics of qubits is caused by the dynamical Zeeman and Bloch-Siegert effects resulting in the shift $\delta\Omega$ of the Rabi oscillations observed in the MW field. The shift induced only by the dynamical Zeeman effect depends on the sign of detuning, $\omega_{rf} - \omega_1$. This fact allows us to separate the contributions of the dynamical Zeeman and Bloch-Siegert effects. In the dispersive limit, it is also possible to take analytically into account dissipative and dephasing processes by introducing the standard relaxation operator in the Liouville equation for the density matrix of the system under study. The Rabi oscillations in both the MW and the bichromatic fields decay at the same rate. In the doubly rotating frame, the qubit's population difference in the bichromatic field tends to the steady state undergoing damped oscillations, whereas without the RF field the population difference decreases exponentially. Results of our experiments are in a very good agreement with theoretical predictions. The model of interaction of qubits with the bichromatic field in the dispersive regime can be effectively used in the Rabi spectroscopy. Within the framework of such an approach, spectroscopic and relaxation characteristics of matter as well as the parameters describing the interaction of qubits with external fields can be studied. The proposed model is also suitable for investigations of such fundamental phenomena as the dynamical Zeeman and Bloch-Siegert effects.

**Figure legends**

FIG. 1. The Rabi oscillations of the $E_1'$ centers in crystalline quartz for different values of the RF field detuning. (a) The signals detected at $\omega_{mw} = \omega_0$ and $\omega_1/2\pi$ = 1.03 MHz: for the black line $\omega_2 = 0$, for the short dot line $\omega_2/2\pi$ = 0.24 MHz, $\omega_{rf}/2\pi$ = 0.20 MHz, for the gray line $\omega_2/2\pi$ = 0.24 MHz, $\omega_{rf}/2\pi$ = 2.0 MHz. (b) Results of analytical calculation based upon Eq. (13) for the parameters used in the experiment, with color coding of lines as in (a).

FIG. 2. The effective Rabi frequency vs the frequency of the RF field at different RF amplitudes: $\omega_2/2\pi$ = 0.12 MHz (squares), 0.24 MHz (triangles), 0.31 MHz (black circles),

0.44 MHz (white circles). $\omega_{mw} = \omega_0$ and $\omega_1/2\pi$ = 1.03 MHz. Solid curves are calculated by using formula (12).

FIG. 3. The frequency shift of the Rabi oscillations vs the frequency of the RF field at $\omega_1/2\pi$ = 1.03 MHz and $\omega_2/2\pi$ = 0.24 MHz: The black line represents the total shift which is the sum of the shifts due to the dynamical Zeeman effect (gray line) and Bloch-Siegert effect (dashed line).

FIG. 4. The calculated time evolution of the population difference with (the black line) and without (the gray line) the RF field in the singly (a) and doubly rotating (b) frames. The signals were obtained from Eq. (11) with the following parameters: $\omega_1/2\pi$ =1.05 MHz, $\omega_2/2\pi$ = 0.24 MHz, $\omega_{rf}/2\pi$ = 0.50 MHz, $T_1$ = 200 μs, $T_2$ = 3.5 μs, $T_2^*$ = 1 μs is the inverse linewidth of the Gauss function $g(\Delta)$, $\gamma_{12}$ =0.

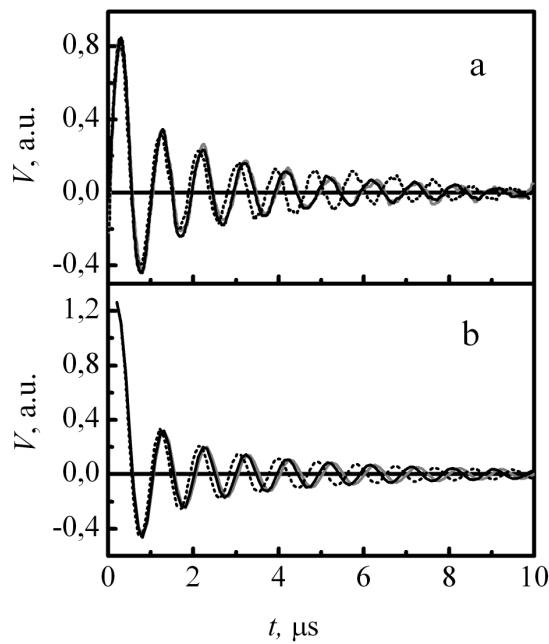

Fig. 1.

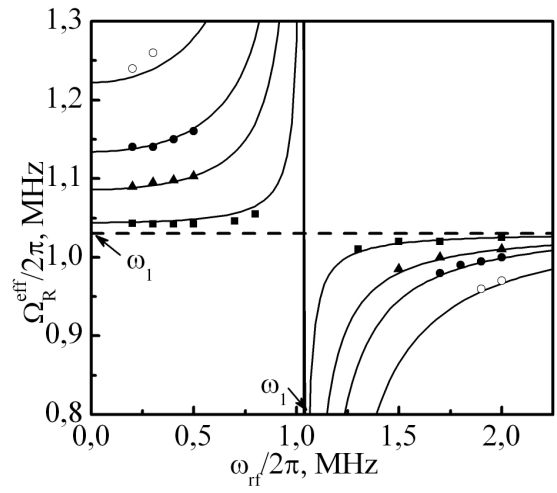

Fig. 2.

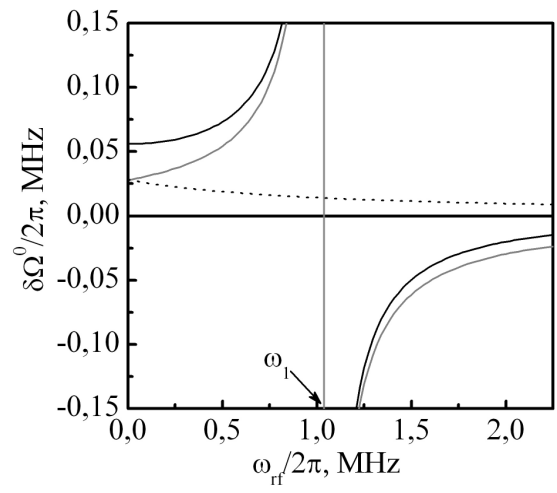

Fig. 3.

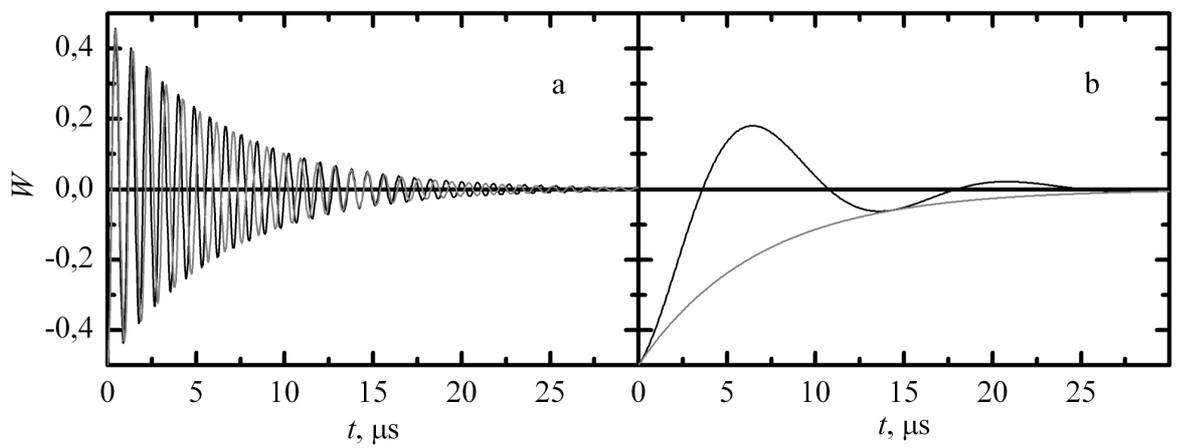

Fig. 4.

11